# A Miniscule Survey on Blockchain Scalability

Angela Li
li.10011@osu.edu
The Ohio State University

Guanqun Song
song.2107@osu.edu
The Ohio State University

Ting Zhu
zhu.3445@osu.edu
The Ohio State University

*Abstract*—With the rise of cryptocurrency and NFTs in the past decade, blockchain technology has been an area of increasing interest to both industry and academic experts. In this paper, we discuss the feasibility of such systems through the lens of scalability. We also briefly dive into the security issues of such systems, as well as some applications, including healthcare, supply chain, and government applications.

## I. INTRODUCTION

Since Bitcoin's conception in 2008 and subsequent launch in 2009, there are now thousands of different cryptocurrencies such as Etherum, Tether, and Dogecoin. Currently, there are hundreds of millions of cryptocurrency users from all around the world. In addition to cryptocurrency, another type of digital asset—non-fungible tokens (NFTs)—were created in 2014 and have been increasingly popular, with millions of unique buyers and sellers today. Moreover, there a billions of US dollars in NFT sales annually, with some NFTs being sold for millions of US dollars. Both cryptocurrency and NFTs are an application of blockchain technology, a type of decentralized, distributed database where users can make direct transactions with other users.

Although blockchain is currently used mainly for cryptocurrency and NFTs, many papers have explored applying blockchain to other domain applications such as healthcare, supply chain, and government identification. Beyond academic papers, there has been action in the industry world. For example, there have been multiple startups such as PokitDok and Avaneer Health, both of which have received tens of millions of US dollars in funding. attempting to apply blockchain to healthcare. One potential application of blockchain in the direction of smart health is in the development of " smart health " solutions such as wearables or connected medical devices to improve existing advanced technologies. By using distributed ledger systems, these devices can securely store and track data, allowing patients to monitor their own health conditions remotely [1-4].

The distributed nature of blockchain has tremendous potential to revolutionize industries other than healthcare. Revolutionize other industries outside of healthcare. For example, in the energy sector, the decentralized nature of blockchain can be used to securely track and manage the distribution and dispatch of energy This can also be combined with the Internet of Things [5-15]. This can ensure that energy is used at the right time to effectively reduce waste and thus increase efficiency.

Nowadays, articles claiming that Web3—the vision of a decentralized Internet built on blockchain—as the future flood the Internet. This buzz around Web3 is due to the idea of users taking power back from large companies such as Google, Amazon, and Meta that dominate the Internet. Many people are worried about the amount of data these large companies have on everyone who uses the Internet, as well as how their ability to censor content and control whose content is most visible. By decentralizing the Internet via blockchain, users get more control over their data because they directly manage their data rather than a third-party managing their data.

Given the rising popularity of blockchain and examinations into its application that could transform entire industries, it is necessary to understand the constraints and feasibility of blockchain. Only by understanding the downsides, can we truly understand the benefits that could come with adopting blockchain for different applications.

This paper will be structured as follows: first, we will give a high level background on blockchain technologies and some of its scalablility and security challenges We will then do a deep analysis on scalability solutions in blockchain, namely sharding solutions.

## II. RELATED WORK

There have been multiple other surveys on blockchain technologies in the past few years. For example, [16-18] provide a survey on blockchain's security issues, while [19-20] provide a survey in the context of different domain applications, namely healthcare and cryptocurrency. [21] analyzes blockchain specifically in the context of supply chain applications. [22-24] examine blockchain in healthcare, and [25] further examines blockchain for government agencies. [19] gives many examples of potential usages of blockchain. For example, blockchain could be used in facilitating transactions involving digital music. In the music industry, there are often copyright disputes—famously, Taylor Swift's disputes over ownership of her past albums that led to her rerecordings of said albums—, so having a permanently and available-to-all database of music ownership could resolve these issues. Another example in [20] is to use blockchain for voting in elections. Currently, in the United States of America, elections on both the federal and state level are done by physical ballots due to worries about the insecurities of electronic voting. To prevent any fraud or tampering in electronic voting, blockchain could keep an irreversible record of every person's vote. Moreover, blockchain could be used for the government by keeping track of government funds similar to how it already keeps tracks of all cryptocurrency transactions. This way, it will be unlikely for any fraudulent behaviour regarding government funds.

Although there are a variety of domain applications, all these applications take advantage of the fact that blockchains are immutable. The data stored in the blockchain, once added, is extremely difficult to modify or delete due to properties we will discuss in the following background section. In healthcare, blockchain is proposed to be used for managing medical records to prevent forgery. Similarly, for government agencies, blockchain is proposed to be used for managing government identification information. Moreover, in the current use of blockchain—cryptocurrency—, blockchain's immutability prevents fraudulent financial transactions [26-27].

Past work surveying blockchain's security issues considers attacks such as exploitation of software bugs in the blockchain—the most prominent example being an exploitation of integer overflow in the Bitcoin software in 2010—, exploitation of smart contract vulnerabilities, exploitation of cryptocurrency wallets, and also low risk attacks like the 51% vulnerability. The 51% vulnerability is an attack where the majority of nodes in the blockchain network are malicious, but it is difficult to achieve such wide control of the network due to cost. The malicious user would have to purchase thousands of machines to take over the network. [16-17] evaluate the state-of-art security software to defend against these attacks such as smart contract bytecode analyzers, network anomaly detection using machine learning, and cryptography techniques like homomorphic encryption, where encrypted data can be directly operated on rather than decrypting the data first. Overall, according to [16-18], there is still a lot of work to do in order to secure smart contracts, which would be a major requirement in adopting blockchain for contexts such as healthcare, where patient privacy is critical.

Past work surveying blockchain's scalability issues note that it is difficult to scale blockchain to the current state-of-art systems without sacrificing decentralization, which is the key point of using blockchain over current systems. For example, in the context of financial transactions, popular cryptocurrencies like Bitcoin can only achieve a throughput of about seven transactions per second due to its proof-of-work consensus protocol, which only approves one block added to the blockchain approximately every ten seconds. However, credit cards like Visa can achieve tens of thousands of transactions per second. Even with works such as [28-29], which can improve cryptocurrency transactions to be comparable to credit cards, there is a lot to be desired in terms of scalability. For cryptocurrency, throughput is improved at the cost of decentralization, as many solutions propose sharding or partitioning the blockchain such that fewer nodes participate in verifying the blocks. Fewer nodes participating in verification may also lead to security vulnerabilities since instead of the 51% vulnerability, a malicious party could take over a smaller portion of the network to take over the verification process. Past work on blockchain's scalability also consider a variety of solutions, grouping them into three main categories based on which layer of the blockchain architecture they affect. There are layer 0 solutions, which target how data is physically transmitted in the blockchain. Next, there are layer 1 solutions, which target the blockchain's protocols such as the data contained in blocks and consensus mechanisms. Finally, in layer 2, there are solutions that target the application lying on top of the blockchain. Such solutions are called off-chain methods, and a notable method is moving some of the computation costs of blockchain consensus mechanisms to the application itself.

In addition to surveying security, scalability, and different applications of blockchain, works such as [20] consider blockchain through other feasibility factors like energy efficiency, ease of integration with current systems and ease of use, and issues with regulation. Infamously, regarding energy efficiency, Bitcoin's proof-of-work consensus mechanism uses a massive amount of computation and thus energy. A different consensus mechanism called proof-of-stake circumvents this problem.

In the future blockchain could also be closely linked to low-power communications, which would both secure Internet of Things (IoT) devices and create distributed wireless networks in smart cities to improve security and efficiency. Leveraging the low-power communications enabled by backscatter [30-32] and the confidentiality of blockchain can transform not only traditional forms of business, but also future communication protocols [33-40]. In addition, we can borrow the decentralized features of blockchain and further research including wireless charging, navigation, concurrent communication cross protocols, etc [41-55].

III. BACKGROUND

In this section, we will give a high level overview of blockchain technologies, especially in context of cryptocurrencies as it is the current main application of blockchain.

Firstly, what is blockchain? As previously described, blockchain can be viewed as a database distributed across all nodes in a P2P network. Although, now that blockchains such as Bitcoin are hundreds of gigabytes large and only increasing in size, only a minority of the nodes in the network actually store the entire blockchain on their machines. An important characteristic of blockchain is that it is decentralized, meaning there is no central authority that controls and verifies the database. One such central authority on today's Internet is Google, which manages all the data of people that use their search engine and other products. Another such authority are banks. Often, to make a transaction today, one must initiate a transaction that is verified and approved by one's bank, who transfers the money to the intended target. In a blockchain world, such as with cryptocurrency, currency is directly transferred from one user to the other without a third-party authority like the bank to verify the transaction. Instead of authorities, transactions are verified by multiple nodes in the network who must all agree—or come to a consensus—about whether to approve or reject the transaction.

The main component of blockchain are blocks, which are just the data stored in the blockchain. In the context of cryptocurrency, blocks often contain the following key components: a time stamp of the time of transaction, a nonce which is a number used in calculating the hash of the block, the hash of the block itself which is usually SHA256 of all the block's contents, the merkle root which is the hash of all blocks on the current chain so it is a summary hash, and a batch of transactions. Most importantly, each block contains the hash of the previous block, meaning that each block points to another block, thus forming a chain of blocks, hence the name blockchain. Since all the block's components are involved in calculating its hash, any change to any of the block's components will lead to an incorrect hash. For example, if any of the transactions in a block were tampered with, then the block's hash will change and thus become inconsistent. This way, the nodes in the network will realize that the block is incorrect. Hence, consistency is maintained through many nodes realizing that a block is incorrect. Since the previous hash is also included in calculating the current hash, any change in a block that makes it incorrect will also make all subsequent blocks incorrect. In more detail, if a block becomes incorrect, then its hash changes, so the block pointing to the first block will become invalid because it is no longer pointing to a valid block. Subsequently, the block pointing to the second block will now be pointing to an invalid block, so that block will also become invalid. And this process of invalidity goes on until the end of the chain, or multiple chains, if the chain branches.

Since there is no central authority to confirm blocks of transactions, the blockchain uses consensus mechanisms to do so. Now, we will introduce two of the main consensus algorithms: Proof-of-Work and Proof-of-Stake.

*A. Proof-of-Work (PoW)*

Proof-of-Work is the consensus algorithm used by Bitcoin, and previously used by Ethereum. In this mechanism, there are miners, which are nodes that take on the task of verifying and adding new blocks to the blockchain. In the Bitcoin blockchain, a block can be added once a miner finds a nonce value that makes the hash value of the new block smaller than a target value. The target value is adjusted based on the current difficulty, which is changed every 2016 blocks so that it takes an average of ten minutes for a new block to be added. By taking ten minutes to add a new block, it would be very difficult and slow for a malicious party to add a bunch of malicious blocks. Unfortunately, this limits throughput to approximately a meager seven transactions per second. The Bitcoin blockchain uses SHA256 hash function to calculate the hash of each block. Due to the random nature of the SHA256 function, the only way for miners to find a nonce value that makes the hash satisfy the target value is by randomly guessing nonce values. Today, it takes approximately $2^{45} = 35184372088832$ computation attempts to find a nonce value that will give a valid hash. In order to produce a block about every ten minutes, miners in the Bitcoin blockchain compute approximately 242 exahashes per second, which is extremely compute intensive. As mentioned in many blockchain papers, the computational intensity of this consensus mechanism in Bitcoin alone has enormous energy implications. In fact, the energy consumption of Bitcoin mining outweighs those of some countries, including Argentina, which has a population of over forty million people. Once a miner finds a good nonce value, then they get to add the block to the blockchain and then, they are rewarded with Bitcoin. The cryptocurrency reward for adding a block incentivizes nodes to participate in mining.

Unfortunately, the large computation power required, which thus requires expensive and beefy hardware, and the large size of the Bitcoin network, which is a few hundred gigabytes, leads to only a minority of nodes on the Bitcoin network to be miners. Miners are a subset of a type of node called full-nodes, which store the entire blockchain. These nodes check the validity of transactions which are then picked up by miners and put into blocks, and then attempted to be completed by adding the block to the blockchain. There is no financial reward for being a full-node, so there are few full-nodes that are not also miners. However, practically all miners are full-nodes because they must verify the transactions before attempting to add a block to the blockchain. If a miner finds a good nonce but then the block actually contains an invalid transaction, then they just wasted all that computation power without getting a Bitcoin reward.

*B. Proof-of-Stake (PoS)*

Proof-of-Stake is the consensus algorithm used by Etherum. In this consensus algorithm, users stake their coins in order to validate transactions. These users are called minters or validators. Staking coins means that the minters are locking up their cryptocurrency in order to create new blocks and confirm transactions. Minters are randomly selected to confirm the next block. In return for staking their coins, users are rewarded with a portion of the transaction fees associated with the block they have validated. This provides an incentive for users to validate transactions, increasing the overall security of the blockchain network.

Proof-of-stake helps to eliminate the negative environmental impacts associated with the energy-intensive process of mining. Additionally, because the nodes are selected randomly, it is more difficult for malicious actors to take control of the network and manipulate the data. Furthermore, proof-of-stake networks are often much more cost-effective than proof-of-work networks because miners are not required to purchase expensive hardware. All of these benefits make proof-of-stake an attractive solution for those looking for an efficient and secure way to validate transactions.

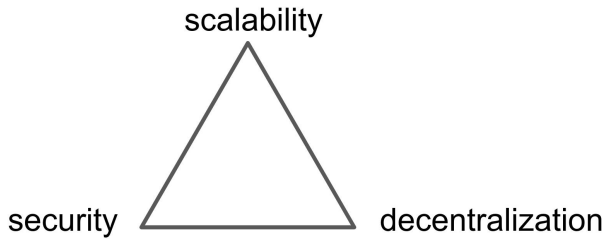
Fig 1.Characteristics of Blockchain

Unfortunately, due to the stake nature of proof-of-stake, this consensus mechanism leads to more centralization as only nodes that already have currency are able to become minters. This centralization problem also has security vulnerabilities. Since nodes with more currency are able to stake more currency, then this consensus protocol is more vulnerable to a monopoly of the network by large stakeholders. This has the potential to create a situation where the interests of the large stakeholders are prioritized over the interests of the average user. Moreover, proof-of-stake is more vulnerable to certain attack vectors such as the nothing-at-stake attack and long-range attack.

In a nothing-at-stake, a malicious actor can take advantage of the fact that there is nothing at stake for them when they sign blocks. This means that they can sign multiple versions of the same block, creating a fork in the blockchain. By doing this, they can double spend their coins and harm the network. They can also manipulate the blockchain by creating a transaction that will only be included in one of the forks, allowing them to take advantage of the fact that the other fork won't be recognized. In a long-range attack, malicious users create a fork in the blockchain where the malicious user controls most of the coins in the fork. This allows them to double spend coins and manipulate the blockchain to their advantage.

In addition to these weaknesses, the proof-of-stake mechanism is more complicated to implement than the proof-of-work mechanism. Therefore, on smaller blockchain networks that have less resources, it would be difficult to implement the proof-of-stake mechanism.

Now, before diving into the general scalability and security issues of blockchain, we will first introduce the scalability-security-decentralization trilemma. This trilemma, as illustrated above, states that only two of the three desired properties can be satisfied simultaneously. Since blockchain is a type of distributed database, this trilemma is a corollary of the CAP theorem, which states that a distributed database can only simultaneously satisfy two of the following three properties: consistent, available, partition tolerance.

*C. Scalability Issues*

Evidently, PoW is not very scalable because a block is only added to the blockchain approximately every 10 minutes. To add more blocks and thus increase transaction throughput, we could lower the difficulty to mine a block. Unfortunately, lowering the difficulty to mine could lead to serious security implications as a malicious user could add their malicious blocks more quickly. In addition, throughput can be harmed with high network congestion. Beyond throughput and latency issues, proof-of-work consumes a lot of energy, and will only increase more energy as the number of miners increase with the size of the network. A third issue that applies to all blockchain networks regardless of consensus protocol is storage. As stated previously, the size of popular blockchain networks is already hundreds of gigabytes large, which can very rarely fit on an ordinary laptop. In order to address this problem, not all nodes have to store all of the blockchain, but this leads to centralization. We saw this problem occur in Bitcoin's blockchain, where the number of full-nodes are in the minority.

*D. Security Issues*

There are a few security issues in blockchain that can occur due to malicious actors, such as 51% attacks, double spending, and blockchain forks. A 51% attack occurs when a malicious party gains control of more than 50% of the network's computing power, allowing them to manipulate the blockchain ledger and double spend coins. Double spending is another security issue in which a malicious party sends the same cryptocurrency to two different addresses, allowing them to spend the same funds twice. Lastly, blockchain forks are security issues that occur when two separate chains form from a single blockchain, resulting in a loss of data and funds.

## IV. SCALABILITY

In this sections, we will examine recent papers addressing the issues of scalability. There are many papers addressing the scalability of blockchain, notably OmniLedger, Monoxide, and OHIE.

*A. OmniLedger*

OmniLedger is a sharding solution that is compatible with both proof-of-work and proof-of-stake consensus algorithms. The main idea of OmniLedger is to partition the blockchain into multiple shards such that each shard can verify blocks in parallel. In order to increase security, OmniLedger randomly assigns and reassigns validators every so often to prevent malicious parties from continuously controlling a shard. Another main contribution of OmniLedger is to ensure atomicity in cross-shard transactions. A huge problem with cross-shard transactions is that if a transaction that sends money is verified in one shard, but the receiver in another shard rejects that transaction, then the currency will be sent but not received. There is also no way to recover the currency because the block is not verified and added to the blockchain, so there is no record of such a transaction occurring. OmniLedger takes an approach seen in traditional parallel databases: locking. Before sending the currency, OmniLedger locks the transaction and only unlocks it after the receiving party also approves the transaction. Unfortunately, locking may increase latency and decrease throughput. To improve throughput, OmniLedger proposes optimistic validation of blocks, which also seems to be inspired by the optimistic method in traditional parallel databases. In OmniLedger, smaller blocks—blocks containing fewer transactions—are first optimistically validated, which leads to lower latency. Since the transactions are small, validating these blocks actually leads to lower throughput than if we validated larger blocks, or larger batches of transactions at the same time. The transactions in the smaller blocks are then batched with more transactions and reverified in a larger block, improving consistency. Unfortunately, a clear consequence of this mechanism is that if a transaction in a smaller block was wrongly approved, then it would have already gone through and would need to be undone, which will increase latency.

OmniLedger was tested on a network of 18,000 nodes, which is not very comparable to the millions of credit card users. The results show that OmniLedger's optimistic validation mechanism greatly improves the system's throughput. When there are few adversaries, for example composing of only 12.5% of less of the blockchain network, OmniLedger is able to surpass 4000 transactions per second, which they claim is higher than Visa. However, OmniLedger's latency per block is rather high, with latency around ten seconds and increasing as the number of nodes in the network increases. Moreover, the paper only evaluates the system with the number of shards ranging from one to seven, which is much fewer than the number of partitions in a traditional relational database that could have at least twenty partitions. The small number of shards corresponds to the small number of nodes that OmniLedger was tested on. Hence, there is still a lot of work to be done on improving scalability.

### B. Monoxide

Monoxide is another scalability solution that uses sharding, which are labeled "consensus-zones." Like OmniLedger, these shards are able to verify and add blocks to the blockchain in parallel, and Monoxide supports both proof-of-stake and proof-of-work consensus algorithms. A key characteristic of Monoxide is that each shard does independent mining or validation of blocks. In stark contrast to OmniLedger, which uses a locking mechanism to ensure consistency in transactiosn involving multiple shards, Monoxide uses an asynchronous consistency protocol that guarantees eventual atomicity. This consistency mechanism for cross-shard validation reduces throughput as we are no longer waiting on locks. However, an asynchronous mechanism could lead to temporary inconsistencies that require rollbacks, which will lead to higher latency. Monoxide claims that these inconsistencies should be relatively uncommon. Finally, a third main contribution of Monoxide is Chu-ko-nu mining, which amplifies the mining power for security. In Chu-ko-nu mining, although each shard has independent mining, miners are able to add blocks in multiple shards by mining a nonce that completes the hash for multiple shards. This nonce will be difference from a nonce that will complete the hash for a subset of those multiple shards or for a single shard. By allowing miners to add blocks to multiple shards with a single mine, Chu-ko-nu incentivizes miners to further compete against each other, which will prevent a malicious miner to easily take over a single shard. A malicious miner will have to compete against both miners in their shard that are mining for that single shard, and also against miners that are in different shards that are mining for a group of shards including the one that the malicious miner is in.

Monoxide was tested on 48,000 nodes, which is much more than OmniLedger but is still not comparable to the large authorities that often have hundreds of thousands of concurrent users at once. As a weakness, Monoxide does not test on networks that have adversaries or malicious users, unlike OmniLedger. Moreover, they report that almost ten percent of blocks are wasted by being successfully mined but not accepted by the blockchain network. This number is likely due to the asynchronous consensus mechanism. If a block is originally approved but then later rejected, this would result in a wasted block. Monoxide tested the number of shards from 1 all the way up to just over 1,000. Based on their results, they claim linear scaling of throughput (transactions per second) from one shard up to over 1,000. However, it would be more impressive to see scaling as the number of nodes increased with the number of shards. Similar to OmniLedger, Monoxide is able to achieve throughput comparable to credit cards, up to 10,000 transactions per second as the size of blocks increased. Unfortunately, Monoxide is also similar to OmniLedger in that the latency is high. To confirm a block, it takes Monoxide approximately 15 seconds on average as the number of shards increases to 1,000. When the number of shards is small, the latency is around 7.5 seconds, which is still extremely high compared to using a credit card today.

### C. OHIE

OHIE is another shard based blockchain scalability solution. In contrast to both OminLedger and Monoxide, OHIE only supports the proof-of-work consensus algorithm. Rather than shard groups of nodes together, OHIE shards by lumping groups of blocks together. There are multiple chains of blocks that can be added to in parallel. Hence, miners in OHIE are able to add a block to any of the parallel chains. When a miner successfully mines a block, then the chain to which the block is added to is determined based on the relative lengths of the chains. OHIE's chain length balancing algorithm does not attempt to have all chains be the same length. Rather, the differences in the chain's lengths are more or less preserved. In fact, OHIE keeps the chain's lengths within a constant factor of each other. This main contribution of OHIE is owed to the total ordering mechanism of blocks across all chains. Total ordering also ensures consistency by preventing double spending. With the ability to exactly say which block and hence, which transaction, happened before another transaction, there is no chance of potentially switching the order and spending the same cryptocurrency twice. Finally, a last main contribution of OHIE is that it claims to be much simpler to implement than the other two algorithms.

OHIE was tested on 12,000 nodes, which is the smallest test network of the three papers. Unfortunately, OHIE did not perform as well in throughput compared to both OmniLedger and Monoxide. Moreover, it had similar high latency problems as in the other two papers.

## VI. CONCLUSION

In conclusion, there is still a long way to go before blockchain can be successfully used in industry applications. Although there are many proposed solutions to improving scalability in blockchain, these solutions have many drawbacks. The main few are that 1) the evaluation of such solutions are only on networks consisting of tens of thousands of nodes, not yet reaching the scale of hundreds of thousands of nodes, which is more comparable with the number of concurrent users on large, authoritative networks today; 2) the latency of transactions is quite high due to the time to verify and add blocks. Despite high latency, the throughput can be high by batching many transactions into a single block.

In addition to scalability issues, there are still security implications that require constant work and attention. Even a single software bug in the Bitcoin blockchain network led to thousands of Bitcoin stolen, which is equivalent to millions of US dollars.

Finally, there is no clear application of blockchain in the various applications mentioned in academic papers. Although there are some startups attempting to use blockchain, the vast majority of the world still uses the centralized, more intuitive (but perhaps more intuitive due to its familiarity) network of web2.

Hence, blockchain may not be so quickly adopted into industries as it was for cryptocurrency. However, that does not take away from the fact that blockchain is a new and exciting technology that is worthy of further research. In the future, blockchain can also leverage its unique characteristics to combine with IoT to maximize the use of protocols by collaborating with various advanced wireless sensors to build advanced IoT architectures [56-67].